\documentclass[aps,twocolumn,noshowpacs,superscriptaddress,floatfix,nofootinbib]{revtex4-1}
\usepackage[table]{xcolor}
\usepackage{quantikz}
\usepackage{amsmath,amssymb}
\usepackage{mathrsfs}
\usepackage{mathtools}
\usepackage[justification=justified]{caption}
\usepackage{xcolor}
\usepackage{verbatim}
\usepackage[normalem]{ulem}
\usepackage[rightcaption]{sidecap}
\usepackage{graphicx}
\usepackage{subfigure}
\usepackage{caption}
\usepackage{multirow}
\usepackage{optidef}
\usepackage{amsthm,epsfig,tikz}
\usepackage{float}
\usepackage{multirow}
\usepackage[colorlinks,citecolor=black,urlcolor=black,linkcolor =black,bookmarks=false,hypertexnames=true]{hyperref} 
\usepackage{relsize}
\usepackage[classfont=sanserif,
langfont=sanserif,
funcfont=sanserif ]{complexity} % http://cse.unl.edu/~cbourke/latex/complexity.pdf
\usepackage{hyperref}
\usepackage{enumitem}
\usepackage{caption}
\usepackage{float}
\makeatletter
\let\newfloat\newfloat@ltx
\makeatother
\usepackage{algorithm}
\usepackage{algcompatible}

\usepackage[classfont=sanserif,
langfont=sanserif,
funcfont=sanserif ]{complexity} % http://cse.unl.edu/~cbourke/latex/complexity.pdf
\newclass{\KP}{KnapsackProb}
\newclass{\CO}{ColOpt}
\newclass{\XORSAT}{XORSAT}
\newclass{\NISQ}{NISQ}
\newclass{\TSP}{TravSalProb}
\newclass{\BinPP}{BinPackProb}
\newclass{\MIS}{MaxIndSet}
\newclass{\MAXCUT}{MAXCUT}

\renewcommand{\part}[2]{\frac{\partial #1}{\partial #2}}

\usepackage{float}
\makeatletter
\let\newfloat\newfloat@ltx
\makeatother

\begin{document}

\title{Predicting Ising Model Performance on Quantum Annealers}
\author{Salvatore Certo}
\affiliation{Deloitte Consulting LLP}

\author{Georgios Korpas}
% \email{georgios.korpas@hsbc.com}
\affiliation{HSBC Lab, Innovation \& Ventures,  HSBC, U.K.}
\affiliation{Dept. of Computer Science, Czech Tech. University in Prague, Czech Republic}

\author{Andrew Vlasic}
% \email{avlasic@deloitte.com}
\affiliation{Deloitte Consulting LLP}

\author{Philip Intallura}
\affiliation{ HSBC Lab, Innovation \& Ventures,  HSBC, U.K.}

\date{\today}

\begin{abstract}
By analyzing the characteristics of hardware-native Ising Models and their performance on current and next generation quantum annealers, we provide a framework for determining the prospect of advantage utilizing adiabatic evolution compared to classical heuristics like simulated annealing.  We conduct Ising Model experiments with coefficients drawn from a variety of different distributions and provide a range for the necessary moments of the distributions that lead to frustration in classical heuristics.  By identifying the relationships between the linear and quadratic terms of the models, analysis can be done {\it a priori} to determine problem instance suitability on annealers.  We then extend these experiments to a prototype of D-Wave's next generation device, showing further performance improvements compared to the current Advantage annealers.
\end{abstract}

\maketitle

\section{Introduction}

In classical optimization frameworks, certain problems present substantial challenges due to the vast energy barriers that delineate local and global minima, especially when these barriers scale in magnitude with the system size. Prominent examples include the Ising Spin Glass model \cite{Nishimori2001}, the Travelling Salesperson Problem \cite{applegate2006traveling}, and the Knapsack problem \cite{kellerer2004knapsack}. The search spaces for these problems escalate in complexity as a function of the input $N$ \cite{10.5555/574848}. For the examples above, the Ising Spin Glass model grows at a rate of $\mathcal{O}(2^N)$, the Travelling Salesperson Problem has a factorial growth of $\mathcal{O}(N!)$, and the Knapsack problem exhibits a pseudo-polynomial time complexity of $\mathcal{O}(nW)$, revealing its (weak) \NP-Hard nature. 
Assuming that $\P\neq \NP$ and the validity of the Unique Games Conjecture \cite{Khot2002},
it is well established that computing arbitrary precision approximations to some \NP-Hard problems is itself a \NP-Hard problem. Indeed, even improving upon the present-best performance ratios of presently available polynomial-time randomized approximation algorithms, for example Max-Cut, is \NP-Hard \cite{Goemans1995}. Regardless, quite often, approximation algorithms can be particularly useful for moderately to large sized instances but might not be sufficient in general. It is very common to use heuristics such as relaxations to semi-definite programs or simulated annealing (SA) as candidate algorithms for such large-scale problems since they offer an alternative way to search for good-quality approximate solutions. The issue is, however, very often such ``traditional approaches'', especially the ones reliant on thermal fluctuations, often fail to surmount these barriers and find $c$-approximate solution, particularly when they are of a macroscopic order even for problems of relative moderate size. 

Quantum annealers (QAs) \cite{Chakrabarti2022} offer a promising avenue towards the search for faster approximations, hopefully matching or surpassing the quality of the solutions, since quantum tunneling and quantum fluctuations can potentially penetrate even substantial barriers, especially if they are not overly thick. Specifically, by judiciously introducing quantum fluctuations, either in tandem with or as a replacement for thermal fluctuations, systems that previously exhibited frustration can be guided toward ergodic behavior \cite{Mukherjee2022}, exploring all accessible microscopic configurations over time. To harness this behavior for optimization, one must gradually diminish the quantum fluctuations until they disappear, ensuring convergence to the desired ground state. As such, quantum annealing, holds significant promise for outperforming classical counterparts in certain complex optimization landscapes and this has partly been argued in several works \cite{doi:10.1126/science.aat2025,denchev2016computational,albash2018demonstration,Li2018,streif2019solving,stollenwerk2019quantum,King2021,tasseff2022emerging,yan2022analytical,arai2023quantum,Sampei2023} in the context of quadratic unconstrained binary optimization (QUBO) problems. See \cite{yarkoni2022quantum} for a survey of practical applications and use-cases.

% see https://arxiv.org/abs/2305.06631 [arai2023quantum] 

On the other hand, there are applications which QA has been shown to perform poorly and there have been several studies that claim the inefficiency of quantum annealing as an approach to (approximately) solve large \NP-Hard problem instances \cite{Altshuler2010, MartinMayor2015,Vert2021, Farhi2012PerformanceOT, Hen2011ExponentialCO} with evidence that the problem's complexity has the largest effect \cite{sinno2023performance}. While the limitations of quantum annealing should not come as a surprise given they perform randomized search, another obstacle, much like in the case of gate-based quantum computers, is that errors accumulate within quantum annealers and error mitigation schemes are required towards the search for quantum advantage \cite{Pearson2019}.

However, the motivation for this manuscript derives from previous work in \cite{tasseff2022emerging} where the authors utilized hardware-native problems, denoted as ``corrupted bias ferro-magnets'' or CBFMs.  These problems assign linear and quadratic biases according to some discretized probability distribution given the available nodes and edges available in the graph topology of D-Wave's Advantage system. In \cite{tasseff2022emerging}, the authors note that there is an expectation of more challenging classes of problems to run on this system, which we investigate numerically.

In this paper, by examining performance of QA when attributes of the problem are drawn from different distributions, we devise a scheme which helps to determine which problems, and specifically which distributions, lead to either problems amenable for annealers or conversely problems that annealers are likely to give sub-optimal results. For a basis, SA algorithm with default settings (as provided by the \texttt{neal} Python package) on a AMD Ryzen 7 PRO CPU with 8 cores and 16GB RAM is used as a benchmark. We then extend our experiments to a new prototype quantum annealer, showing some level of increased stability and solution quality in the hardest hardware-native problems. In Sec. \ref{sect:qa} we review the basics of quantum annealing as well as some of its applications. In Sec. \ref{sec:hardware} we provide details on the nature of ``hardware-native problems'' which are of interest to us. In Sec. \ref{sec:hardness} we define a quantitative measure on the hardness of a given problem based on the distribution characteristics mapped to the interaction and magnetic field terms of the Ising Hamiltonian, the ``hardness ratio''. In Sec. \ref{sec:effects} we describe the effects of the hardness ratio on the performance of the quantum annealer. In Sec.\ref{sec:predict} we describe a schema to predict the performance of the annealer using the Knapsack problem.  In \ref{sec:prototype} we discuss the prototype Advantage2 annealer and perform a preliminary benchmarking against the previous system.

\section{Background} \label{sect:qa}
Quantum annealers are for a class of quantum processors specifically designed to solve combinatorial optimization problems of both academic \cite{djidjev2018efficient,Boothby2015,Hernandez2017,kumar2018quantum,Harris2018,King2018} and industrial interest from material science to finance \cite{Rosenberg2016,Neukart2017,Ikeda2019,Venturelli2019,Borowski2020, Feld2019, Kitai2020,Grant2021,Yu2021, phillipson2021portfolio,giron2023approaching,Sampei2023}, with exciting applications in other areas such as unsupervised learning and training classical AI models \cite{schuman2019classical,neven2012qboost}.
%Quantum computers are theorized to solve problems currently intractable on classical computers, potentially offering exponential speedups to some of the hardest problems. %Quantum annealers are processors specifically designed to solve combinatorial optimization problems of both academic \cite{Boothby2015,Hernandez2017,Harris2018,King2018} and industrial insterest \cite{Neukart2017,Ikeda2019, Borowski2020, Feld2019, Kitai2020, Yu2021, giron2023approaching}, of which there are exciting applications in similar areas, especially in areas like unsupervised learning and training classical AI models \cite{schuman2019classical,neven2012qboost}.

Quantum annealing is a type of Adiabatic Quantum Computing (AQC) where an initial multi-qubit quantum state is prepared, aligning with the ground state of an elementary Hamiltonian. This system then undergoes an adiabatic evolution, transitioning towards a final Hamiltonian. The ground state of this final Hamiltonian encapsulates the solution to the target optimization problem. Theoretically, it is established that AQC holds polynomial equivalence to conventional gate-based quantum computing because any quantum circuit can be depicted as a time-variant Hamiltonian with a maximum overhead that is polynomial in nature \cite{Aharonov2008}. However this equivalence breaks when one considers stoquastic Hamiltonians, which is the case with QA. 

Therefore, quantum annealing operates under the AQC framework and allows for transitioning from an initial Hamiltonian with a known ground state to a problem Hamiltonian, with the hope that the system ends up in the ground state of the latter. As elucidated in the seminal work by \cite{farhi2000quantum}, if one considers a quantum system that initiates in the ground state of a Hamiltonian, denoted $\mathcal{H}_b$, and culminates in the ground state of a problem-specific Hamiltonian, $\mathcal{H}_p $, then the state of the system at an intermediate time $t$ within the total operational duration $T$ can be represented as
\begin{align}
    \mathcal{H}(t) = \big( 1-u(t) \big)\mathcal{H}_b + u(t)\mathcal{H}_p,
\end{align}
with $u(t)$ defined as the ratio $t/T$. The problem Hamiltonian is a stoquastic 2-local Hamiltonian, the one-dimensional Ising Hamiltonian \cite{Ruelle1999}, of the form
\begin{align}\label{eq:ising}
    \mathcal{H}_p = \sum^n_{i=1} h_i \sigma^i_z + \sum^n_{i=1}\sum^n_{j=1} J_{i,j} \sigma^i_z \sigma^j_z,
\end{align} where $\sigma^i_z$ is the Pauli-$Z$ operator defined as ${\rm diag}(1,-1)$ for the $i^{\rm th}$ spin, $h_i$ is the strength of the ``magnetic field'' acting on the $i^{\rm th}$ spin of the qubit, and $J_{i,j}$ is coupling strength between spins $i$ and $j$ of the respective qubits. The spins take values in $\{-1,1\}$, and the final solution of this system, with a high-probability, will be the ground state, or state with the lowest energy. 

The interest in QA lies in the fact that many combinatorial optimization problems of interest can be recast as quadratic unconstrained binary optimization (QUBO) problems \cite{Kochenberger2014, glover2018tutorial, yarkoni2022quantum} which, by an easy change of variables, they can be mapped to the Ising Hamiltonian \eqref{eq:ising} and thus hoped to that hard instances can be solved.

%Since the vast majority of  of these problems are known to be \NP-Hard \cite{robinson2013introduction}. Several difficult problems, such as the Traveling Salesperson Problem, have been known for decades. Surprisingly, there have been applications outside of well-known operations research problems, such as neural networks, which may be slightly reformulated as a QUBO optimization problem \cite{sasdelli2021quantum} to solve for the trainable parameters. 

\subsubsection*{Available Quantum Annealing Devices}

Currently, cloud-available quantum annealers are manufactured by D-Wave Systems and controlled through the \texttt{Ocean} SDK They correspond to programmable arrays of qubits that offer promising potential to decrease the time to solution of previously discussed \NP-Hard combinatorial optimization problems. Noise and  qubit topology, the graphical structure of the qubits (nodes) as well as possible connections between them (edges), are the current limitations.  D-Wave's recently retired 2000Q machine featured a simple topology of qubits, denoted as the Chimera topology. The Chimera topology consists of \textit{internal couplers}, where connect pairs of orthogonal qubits and each qubit has degree six. The there are \textit{external couplers} that connects qubits of different internal couplers that are in the same row or column. Finally, the Chimera topology has a recurring horizontal structure of four qubits to four vertical qubits, which one may see is a $K_{4,4}$ bipartite graph.  

%Currently cloud-available quantum annealers are the ones manufactured by D-Wave Systems. Controlled through the \texttt{Ocean} SDK, the are programmable arrays of qubits that offer promising potential to decrease the time to solution of \NP-Hard combinatorial optimization problems discussed previously. Noise and  qubit topology, the graphical structure of the qubits (nodes) as well as possible connections between them (edges), are the current limitations.  D-Wave's recently retired 2000Q machine featured the ``Chimera'' topology, where each qubit has degree 6 (coupled to 6 other qubits). {\color{red}[DISCUSS MORE ABOUT THE TOPOLOGIES]}

\begin{figure}
    \centering
    {\includegraphics[scale=0.3]{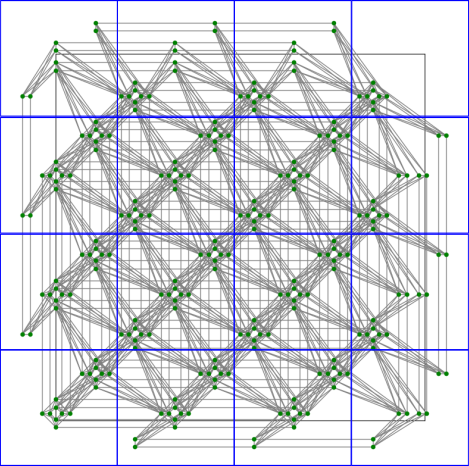}} 
    \caption{The Pegasus topology of D-Wave's Advantage QPUs, shown here with 16 unit cells \cite{boothby2020nextgeneration}.}
    \label{fig:results2}
\end{figure}
The current topology of D-Wave's Advantage annealer is called Pegasus and features higher connectivity than its predecessor, where each qubit has degree 15.  The new prototype Advantage2 system comes with the so-called Zephyr topology that has increased connectivity where each qubit has degree 20, among other enhancements.  These devices can be used for all problem types, even fully-connected problems, by utilizing both minor embedding techniques \cite{Choi2008,Choi2010,Klymko2013,cai2014practical} and decomposition algorithms \cite{Ajagekar2020,bass2020optimizing} available in D-Wave's hybrid solvers at the cost of requiring more physical qubits or possible sub-optimality related to the decomposition \cite{Pelofske2020}.

\section{Hardware Native Problems}\label{sec:hardware}

The incongruity of a problem's graphical structure compared to the QPU's native topology creates overhead such as costly SWAP gates (on gate-based machines) or embeddings with chains of physical qubits to represent one logical qubit.  Because of this, many researchers conducting QA benchmarking choose to utilize problems whose graphs match precisely with the device's topology, referred to as \emph{hardware native problems}.

Given an undirected graph $\mathcal{G} = (\mathcal{N},\mathcal{E})$, with nodes $i \in \mathcal{N}$ and edges $(i, j) \in \mathcal{E}$, define the available qubit biases as $h=\{h_{i}\}, \, \forall \: i \in \mathcal{N}$, and couplings $J= \{J_{i, j}\}, \, \forall \: (i,j) \in \mathcal{E}$.

Having $h,J$ the question is what values should they ideally have so as the problem can be solved on a QA.  Ref. \cite{tasseff2022emerging} proposed a class of hardware native problems, CFBMs, by sampling these coefficients from discrete distributions as shown in the Table below for all $i \in \mathcal{N}$.

%\begin{equation}
%\begin{aligned}
%    P(J_{i,j} = 0) &= .35 \\
%    P(J_{i,j} = -1) &= .10, \\
%    P(J_{i,j}=+1) &= .55 
%\end{aligned}
%\end{equation}
%for all $(i,j)\in \mathcal{E}$, and
%\begin{equation}
%\begin{aligned}
%    P(h_{i}=0) &=.15, \\
%    P(h_{i}=-1) &=.85, \\
%    P(h_{i}=+1)&=0,
%\end{aligned}
%\end{equation}
\[
\begin{array}{|c|c|c|}
\hline 
\textbf{Value } & \textbf{Probability for } J_{i,j} & \textbf{Probability for } h_{i} \\ \hline \hline
 0 & 0.35 & 0.15 \\ \hline
 -1 & 0.10 & 0.85 \\ \hline
 +1 & 0.55 & 0 \\ \hline
\end{array}
\]

This format allows for a direct comparison of the probabilities of  $J$ and  $h$ for each value.

%\begin{equation}
%\begin{split}
%P(h_{i}=0)=.15, P(h_{i}=-1)=.85, \\
%P(h_{i}=1)=0, \ \forall i \in \mathcal{N}
%\end{split}
%\end{equation}
%for all $i \in \mathcal{N}$.

The results from benchmarking these hardware native CFBMs against classical algorithms like SA and Integer Linear Programs (ILP) showed promising and favorable results for D-Wave's QPUs.  Still, the authors noted in \cite{tasseff2022emerging} that further analysis on the problems is needed and that there perhaps exist more challenging hardware native instances on the Pegasus topology.  

We have found that, when the coefficients are drawn from truncated continuous distributions to fit within the hardware energy ranges, we can define a quantitative {\it hardness ratio} that accurately describes the difficulty of the problem for classical algorithms (and subsequent possible advantage for the QPU).

To measure the quality of solutions obtained by QA over its classical counterpart SA, we define the following difference metric:

\begin{equation}\label{eq:rel-dif-sa}
\text{Relative Difference (QA/SA)} \coloneqq \frac{E_{\rm QA} - \min_{E_{\rm SA}}}{|\min_{E_{\rm SA}}|},
\end{equation}
or RL,
where $E_{\rm QA}$ are the energy values obtained from sampling the QA, and $\min{E_{\rm SA}}$ is the lowest energy value obtained from the samples of the SA algorithm.

\begin{figure}[!htb]
    \centering
    \includegraphics[width = \columnwidth]{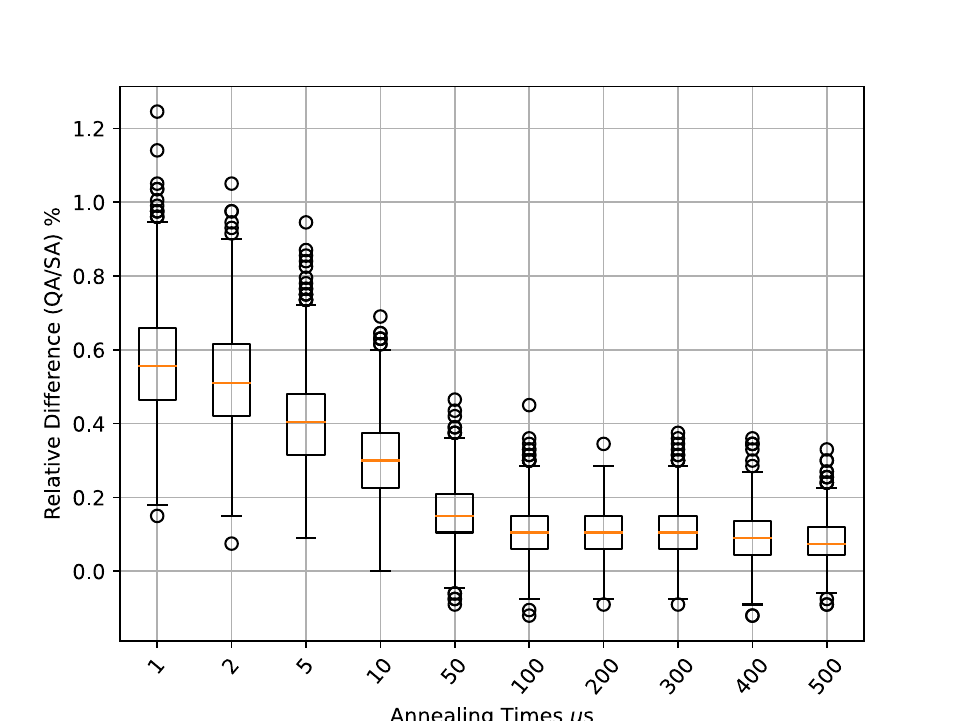}
    \caption{\textbf{Relative Difference from Best SA Solution (RL)}. Boxplots for the difference between the QA samples compared against the lowest energy solution found by the SA.  Using the same CBFM-P instance, we see that using our metric from Eq. \eqref{eq:rel-dif-sa} the QA on average finds worse solutions than the SA.}
    \label{fig:rel-dif-sa}
\end{figure}

Fig. \ref{fig:rel-dif-sa} shows on average the QA is outperformed by SA for the given CBFM instance. The question then becomes what problems does the QA truly have a potential advantage utilizing this direct metric.

Therefore while the CBFM-P instances are difficult for the SA algorithm to achieve consistently low energy solutions, they are not inherently difficult to find a lower energy solution than the average QA solution, as shown in the boxplots of Fig. \ref{fig:rel-dif-sa}.  Still, for most of the longer annealing times, the QA did find some samples that outperformed SA by 0.1\%-0.2\%.  This raises the question of which instances of  Ising Models are more difficult for the SA, and which instances provide the best probability for QA devices to reach low energy solutions the SA cannot, even if the difference is relatively small. These instances are described in the next section.

\section{Hardness Ratio for QUBOs}\label{sec:hardness}

Let us now define new instances of Ising Models which are more difficult for the SA to find lower energy solutions than QA, and therefore lead to a framework for analyzing problems that provide indications of which are most suitable for quantum advantage using QAs.  

Assume that the linear terms are independent and identically distributed (i.i.d) and the quadratic terms are i.i.d., where $\sigma_{h}$ and $\sigma_{J}$ are the standard deviation of the linear and quadratic coefficients of the Ising Model, respectively. If the coefficients are random variables from a continuous distribution with $\mu = 0$, we can define a ratio that quantifies the level of frustration of the particular Ising model:
\begin{equation}\label{eq:hardness}
\mathcal{F} \coloneqq \frac{\sigma_{h}}{\sigma_{J}}.
\end{equation}
Therefore, $\mathcal{F}$ conveys a measure of how dispersed the linear coefficients are in comparison to the quadratic ones.  Intuitively, a problem is difficult for a QA to have advantage over classical algorithms if no quadratic interaction terms exist.  If only linear biases are present, there would be no need for QA as it results to a trivial linear unconstrained problem.  Conversely, if only quadratic terms are present, the combinatorial explosion of the interactions create enormous complexity where classical solver can struggle. As the hardness ratio $\mathcal{F}$ decreases, the implication is that QA could become increasingly advantageous. The location of the inflection point then is of particular importance since it can provide threshold or boundary of a set of indications of which technology to leverage for a given problem.

For D-Wave Systems' QA, the devices on which we conduct experiments, the standard available energy ranges for both the linear $h_{i}$ biases and quadratic terms $J_{i,j}$ are $[-4,4]$ and $[-1,1]$ respectively.  Utilizing uniform continuous distributions, we keep the moments of the quadratic coefficient fixed (uniform distribution with range $[-1,1]$ and $\mu = 0$), so that the hardness ratio can be viewed as the factor difference between the dispersion of the two coefficients.  For example, a hardness ratio of $4$ indicates that the linear coefficients are evenly spread across the entire available $h_{\rm range}$ of $[-4,4]$.

\section{Effects of Hardness Ratio on QA Performance}\label{sec:effects}

\begin{figure}
    \centering
    \textbf{Relative Difference on Hardness Ratios}\par\medskip
    \includegraphics[width = \columnwidth]{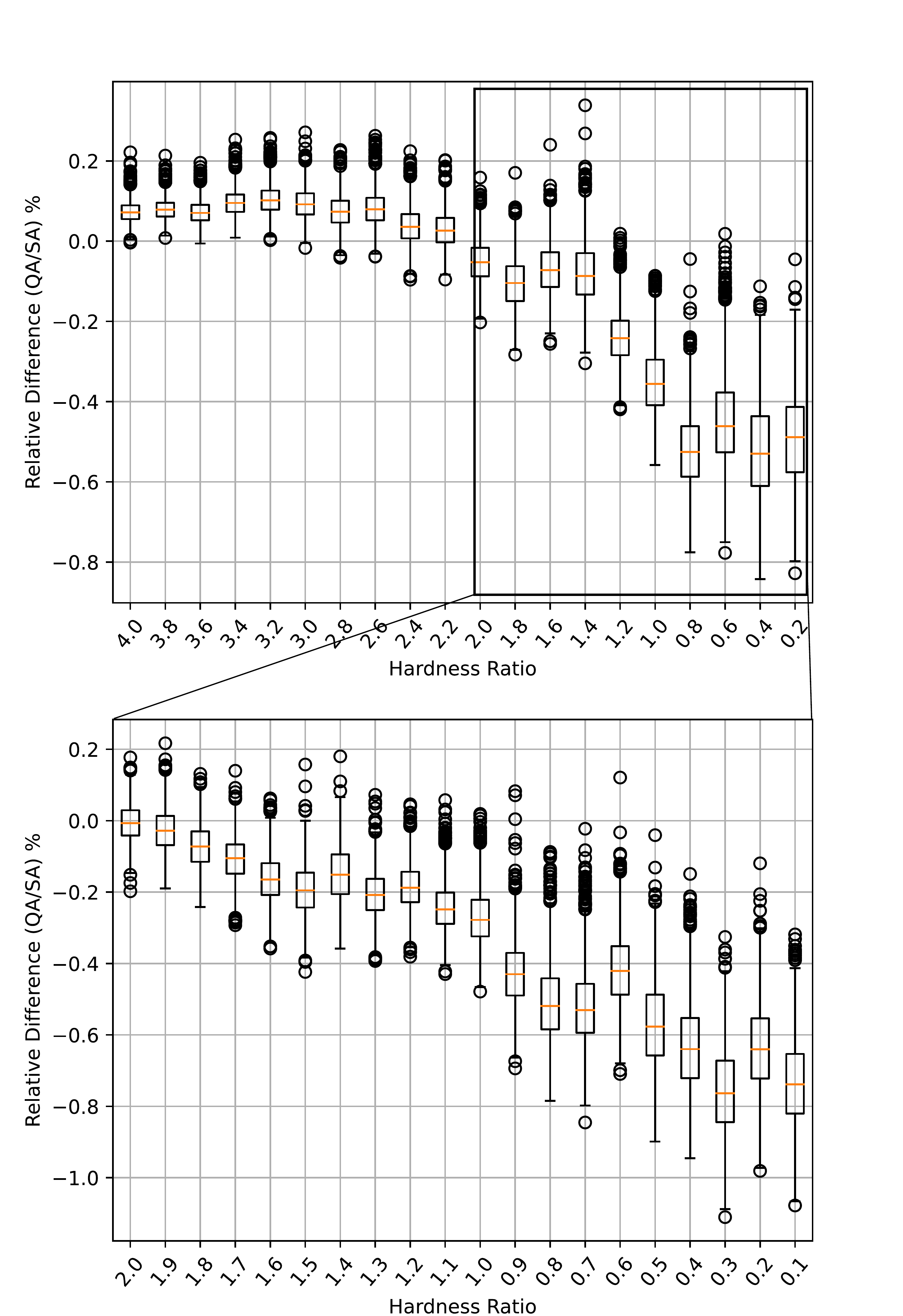}
    \caption{Boxplot of the RL metric as a function of $\mathcal{F}$.  Experiments were conducted on D-Wave's Advantage 4.1 QPU.  For $\mathcal{F} > 2$, the QA shows no advantage on hardware native problems, while for $\mathcal{F} < 2$ the QA has significantly better energy solutions.}
    \label{fig:results5}
\end{figure}

To determine the effects of the hardness ratio on hardware-native Ising problems, we sampled from various uniform distributions across the linear and quadratic energy ranges and compared the quality of the solutions to SA using the RL metric.  Using an annealing time of 100 $\mu s$ and sampling 2000 times each, the RL boxplots are in Figure \ref{fig:results5}. It is clearly evident that a decreasing $\mathcal{F}$ increases the effectiveness of the QA over the SA.  For values of $\mathcal{F}$ above 2 the linear biases dominate over the quadratic terms and the SA consistently outperformed the QA.  A cutoff for relative performance advantage with the QA appears around $\mathcal{F} = 2$ where, below this, the QA has a definitive better performance, consistently averaging better solutions than the best SA solution.

We can view these effects as a function of the reliance on the interactions between the spin variables in driving the classical difficulty of the problem.  When the values of the quadratic terms outweigh the linear, that is, for a low hardness ratio, the classical algorithms have difficulty outperforming QA. Conversely, when the hardness ratio is very high, it is easier for the classical algorithms to find better quality answers as these low energy solutions are dominated by the linear biases.

\section{Predicting Performance with Hardness Ratio}\label{sec:predict}

In this section, we provide the main result of this work which is a system for predicting the performance of a quantum annealer by using the hardness ratio of Eq. \eqref{eq:hardness}. We will concentrate on the Knapsack (\KP) problem. The \KP~is a widely studied weak \NP-Hard problem where the objective is to find the optimal combination of $N$ items to maximize profit subject to a weight constraint. Although it may not be the best problem (similar to \MAXCUT) for such an approach given that there exists a \FPTAS~for it \cite{Donguk2015,chan:OASIcs:2018:8299,jin2019improved}, \KP~is simple enough to explain and relevant to a number of industrial problems, from resource management in software as a service \cite{Aisopos2013} to a toy-simplification of collateral optimization in finance \cite{giron2023approaching}. As such, \KP~is ideal for our purpose to serve as a prototype to demonstrate the implications of the hardness ratio. 

Specifically, in the $N$-item \KP~each item $i \in [N]$ has a profit value $p_i \in \mathbb{N}$ associated with it, as well as a weight $w_i \in\mathbb{N}$.  The integer program formulation of the problem is: 
\begin{argmaxi}
{x \in \{0,1\}^N}{\sum_{i=1}^N x_{i}p_{i} }
{}{}
\addConstraint{\sum_{i=1}^N x_{i}w_{i} \leq C }
\end{argmaxi}
Here $x_{i} \in \{0,1\}$ is a binary variable indicating whether or not an item is selected and $C \in \mathbb{R}$ is the total (weight) capacity of the knapsack, limiting how many items one can choose depending on their weight
%$p_{i}$ is the profit of the item, $w_{i}$ is the weight, and $C$ is the total capacity of the knapsack, limiting how many items one can choose depending on their weight. 
To solve this problem via a QA, it needs to be transformed to a QUBO. A standard formulation is as follows:

\begin{argmini} 
{x\in \{0,1\}^N}{-\sum_{i=1}^N x_{i}p_{i} + \lambda \left(\sum_{i=1}^N x_{i}w_{i} - C \right)^2 }
{}{}\label{prob:knapsack} 
\end{argmini}
where $\lambda \in \mathbb{R}$ is a Lagrange multiplier enforcing the capacity constraint.  As reported before in \cite{pusey2020adiabatic}, QA fails to efficiently solve the \KP~compared to classical methods.  However, the problem in a QUBO reformulation allows us to analyze how the hardness ratio changes as the distribution of weights and capacity change, which, in turn, allows us to quantitatively support the case of why \KP~might not a good candidate problem for QA.

\subsection*{Analysis of the QUBO Coefficient Scaling for the \KP}

%To confirm the claims above, we observe that the linear coefficients of the quadratic constraint of the \KP~scale faster than the quadratic coefficients as a function of the capacity $C$. 

To corroborate further the aforementioned assertions, it is observed that the linear coefficients associated with the quadratic constraint of the \KP~exhibit a more rapid scaling in comparison to the quadratic coefficients, when considered as a function of the capacity $C$. In general, for $i\in [N]$, the range of the quadratic coefficients $r_{(J)}$ with a given \KP~instance are
\begin{align*}
r_{(J)}  = [\min_i w_i^2, \max_i w_i^2],
\end{align*}
and the range $r_{(h)}$ of the linear terms are within the range
\begin{align*}
r_{(h)} = [C \min_i w_i, C \max_i w_i)],
\end{align*}
where ${\rm min}_i w_i$ (${\rm max}_i w_i$) denotes the minimum (maximum) element in the set of weights $\{w_i\}_{i\in [N]}$.
Therefore, the range of the quadratic terms will remain constant, while the range of the linear terms will be $C ({\rm max}_iw_i - {\rm min}_i w_i) $, linearly scaling with $C$ with a slope defined as $({\rm max}_iw_i - {\rm min}_i w_i)$.

Now, consider the case of $C \geq \max_i w_i + \min_i w_i$. Starting from the assumption, observe that
\begin{equation}\label{eq:max-min-C}
    \begin{split}
        & C \geq \max_i w_i + \min_i w_i \\
       %\iff & C  \geq (\max_w + \min_w) \cdot \frac{\max_w - \min_w}{\max_w - \min_w}\\
       %\iff & C  \geq \frac{(\max_w)^2 - (\min_w)^2}{\max_w - \min_w} \\
       %\iff & C \left(\max_w - \min_w) \right ) \geq (\max_w)^2 - (\min_w)^2 \\
       \Rightarrow \,& C  \max_i w_i - C\min_i w_i \geq \max_i w_i^2 - \min_i w_i^2 \\
       \Rightarrow \, & r_{(h)}(C) \geq r_{(J)}.
    \end{split}
\end{equation}

Let us illustrate the above with a simple example. Assume a \KP~instance with $w_i \sim \mathcal{N}(50, 15^2)$, where $\sigma=15$ is the standard deviation, and $\max_i w_i \leq C \leq \sum_{i} w_{i}$.  Assume $\max_i w_i = 90$ and $\sum_{i} w_{i} = 5075$, then using \ref{eq:max-min-C}, if $C\geq 106$ then the range of the linear terms will be greater than the quadratic terms and the hardness ratio will scale linearly as $C$ increases, as we discussed earlier.

\begin{figure}
    \centering
    %\textbf{Hardness Ratios for KP Instances}\par\medskip
    %\def\svgwidth{\columnwidth}
    \includegraphics[width = \columnwidth]{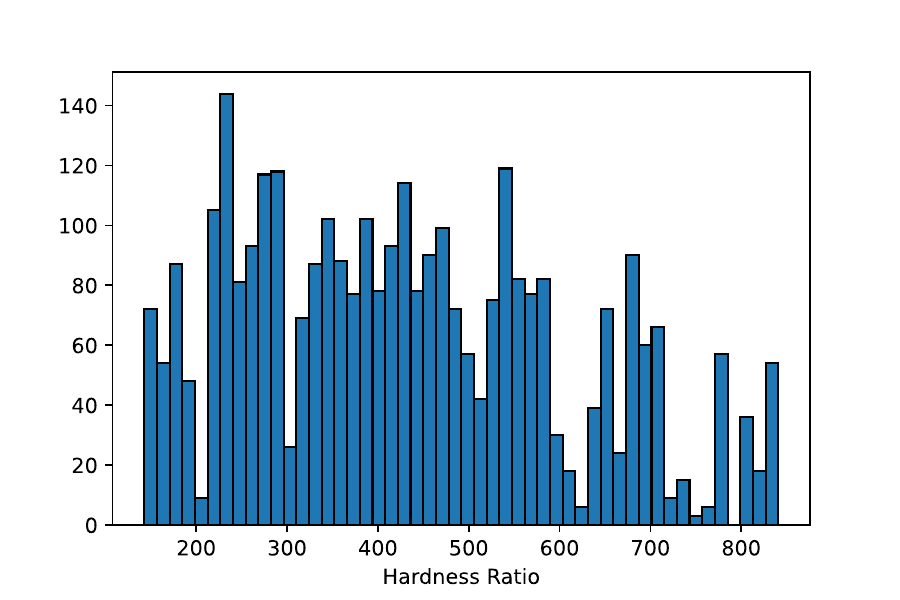}
    \caption{Histogram of the hardness ratios for 3,240 \KP~instances from \cite{Jooken:2022}.}
    \label{fig:KP-Hist}
\end{figure}

Of course, it is a good assumption that $\min_i w_i < C < \max_i w_i$, which implies $C$ is dependent on the given values of the weights. Adjusting for the randomness of the quantities and dependence of the capacity, we take the assumption that the weights are i.i.d. and independent of $C$. Note that if $C< \min_{w}$ then clearly the problem is infeasible and if $C \geq \sum_{i} w_i$ then the solution is trivial. Hence, we will only analyze observations where $C$ is between these values. Therefore, with realistic values of $C$, the hardness ratio will be large.  For example, we considered 3,240 \KP~instances from \cite{Jooken:2022} and transformed them into QUBOs as defined in Prob. \eqref{prob:knapsack}.  As shown in Fig. \ref{fig:KP-Hist}, the hardness ratios are predominantly larger than 1 for all of these instances.

\subsection*{Statistical Analysis of Maximum-Minimum Differences in Random Variables}

To dive deeper into the analysis, we consider arbitrary distributions for the i.i.d. random variables $\{w_i\}_{i\in[N]}$ and for the random variable $C$ in order to understand how the probability of the lengths of $r_{(J)}$ and $r_{(h)}$ affect the problem. In particular, we will derive analytic closed-form expressions for the c.d.f. of the random variable $(\max_iw_i)^2 - (\min_iw_i)^2$ and for the random variable $C(\max_iw_i) - C(\min_iw_i)$.

The following derivations can be found in App. \ref{Sec:dist-der}. For arbitrary random variables $\{X_i\}_{i\in [N]}$, $N\in \mathbb{N}$, define $\mbox{max} =\max\{X_1,\ldots,X_n\}$ and $\mbox{min}=\min\{X_1,\ldots,X_n\}$, the c.d.f. of the difference has the form % = probability of the difference between the maximum and minimum weights being less than or equal to a certain value x
\begin{equation}\label{eq:diff-deriv-FINAL}
   P\big( \max - \min \leq x \big) = \int_{0}^{\infty} \int_{y}^{y+x} p(y,z) p_{\min}(y) dz dy,
\end{equation}
and hence the p.d.f. has the form
\begin{equation}\label{eq:max-min-pdf}
    p_{\max-\min}(x) = \int_0^{\infty} p(y,y+x) p_{\min}(y) dy.
\end{equation}
With Eqs. \eqref{eq:diff-deriv-FINAL}, \eqref{eq:max-min-pdf} and with the c.d.f. of $\{w_i\}_{i\in[N]}$ denoted as $P_w$ and the corresponding p.d.f. as $p_w$,  we next to derive closed form solutions for the \KP. For $r_{(h)}$ observe that $p\Big( C\max_i w_i - C\min_i w_i \Big) = p\Big( C(\max_i w_i - \min_i w_i) \Big)$. For convenience we set $\max \equiv \max_i X_i$ and $\min \equiv \min_i X_i$, where $X_i = w_i$ or $X_i = w_i^2$ given $r_{(h)}$ or $r_{(J)}$, respectively. From the independence assumption with the random variable $C$, the c.d.f. of this range is determined by the convolution (p.d.f.) (see App. \ref{Sec:dist-der}):
\begin{equation}\label{eq:two-vars}
    p_{C(\max - \min)}(z) = \int_{0}^{\infty} p_{\max-\min}(x)p_c(z/x)\frac{1}{|x|}dx.
\end{equation}
Breaking down the integrands, for the first one, $p_{\max-\min}(x)$, we have
\begin{equation}\label{eq:min-pdf}
    p_{\min}(y) = np_w(y)(1-P_w(y))^{n-1},
\end{equation}
and for the second one
\begin{equation}\label{eq:norm-diff-pdf}
    \begin{split}
     p(y,y+x) & = n(n-1)p_w( y+x ) p_w( y)\\
     & \times \Big( P_w(w \leq y+x) - P_w(w \leq y) \Big)^{n-2}.
    \end{split}
\end{equation}
Plugging Eqs. \eqref{eq:min-pdf} and \eqref{eq:norm-diff-pdf} into Eq. \eqref{eq:max-min-pdf} and taking this equation into Eq. \eqref{eq:two-vars} we obtain the final p.d.f. for the random variable $C(\max - \min)$. The behavior of the final p.d.f  is thus dictated by the p.d.f. of $C$ since the fraction $p_C(z/x)/x$ dramatically changes behavior. When $x$ is close to $0$ the function $p_C(z/x)/x$ becomes larger and increases the width of the tail. Contrary to this increase in width, as $x \to \infty$ the function $p_C(z/x)/x$ tends to $0$ much faster, which decreases the width of the tail.

For $r_{(J)}$ observe that since the weight set $\{w_i\}_{i\in \mathbb{N}}$ is $\mathbb{R}^*$-valued the c.d.f. $P_w(w^2 \leq x) = P_w(w \leq x^{1/2} )$ and, given this, Eq. \eqref{eq:diff-deriv-FINAL} has a similar form as the c.d.f. of $\max_i w_i - \min_i w_i$ except with fractional terms, where
\begin{equation}\label{eq:sqrt-min-pdf}
    p_{\rm min}(y) = \frac{y^{-1/2}}{2} np_w( \sqrt{y} )(1-P_w(\sqrt{y}))^{n-1}
\end{equation}
and
\begin{equation}\label{eq:sqrt-diff-pdf}
    \begin{split}
     p(y,x+y) & = \frac{\big( (x+y)y \big)^{-1/2}}{4} n(n-1) p_w( \sqrt{x+y}) p_w( \sqrt{y}) \\
     & \times \Big( P_w(w \leq \sqrt{x+y}) - P_w(w \leq \sqrt{y}) \Big)^{n-2}
    \end{split}
\end{equation}

Plugging Eqs. \eqref{eq:sqrt-min-pdf} and \eqref{eq:sqrt-diff-pdf} into Eq. \eqref{eq:diff-deriv-FINAL} displays the complexity of the c.d.f. where we see an increase in the depth of the tail when $x$ or $y$ is close to $0$, and a decrease in the width of the tail as $x \to \infty$ or $y \to \infty$. Hence, the behavior of this p.d.f. is similar to the p.d.f. of $C(\max-\min)$, however, with two variables inflating the width of the tail. 

Thus, given the similarities of the p.d.f.s between $\max_i w_i - \min_i w_i$ and $\max_i w_i^2 - \min_i w_i^2$, the behaviors of each p.d.f. are dictated by the fraction terms $p_C(z/x)/x$ and $\displaystyle \frac{y^{-1/2}}{2} \cdot \frac{\big( (x+y)y \big)^{-1/2}}{4}$, respectively. Furthermore, given that the distribution of the weights and capacity are well-behaved (ergo, the mean and variance of the variables $(\max_w)^2 - (\min_w)^2$ and $C(\max_w - \min_w)$ are finite), and taking the assumption that the means are equivalent, the behavior of the tails imply there are very few instances where the inequality $\sigma_h < \sigma_j$ holds. Therefore, a general problem will more than likely yield a hardness ratio greater than $1$.

In fact, even for small $C$ values the hardness ratio tends to be greater than $1$. For realistic problem instances, $C$ will be sufficiently large enough that the hardness ratio will be quite large, allowing this QUBO to be easily solved by classical algorithms like SA and therefore not appropriate for QA.

\section{Advantage2 Prototype}\label{sec:prototype}

Recently, D-Wave Systems have made available a prototype of their new Advantage2 QPU \cite{advantagetwo}, allowing users to test the efficacy of the improvements made on the hardware compared to its predecessors.  The Advantage2 Prototype has 563 working qubits on the new Zephyr topology.  While not all improvements,  such increased energy scales, are available on the prototype device, other benefits derived from the new fabrication process are.  

Due to the different topology and qubit-connectivity, comparing the Advantage2 Prototype to currently available systems like Advantage 4.1 is challenging as the hardware native problems are completely different and the solutions not directly comparable.  Still, we can determine how the new prototype device performs in terms of the consistency of the solutions as a function of annealing time, as well as the consistency based on the specific hardness ratio that we introduced for the Ising Model.

\begin{figure}[!ht]
    \centering
    \includegraphics[width = \columnwidth]{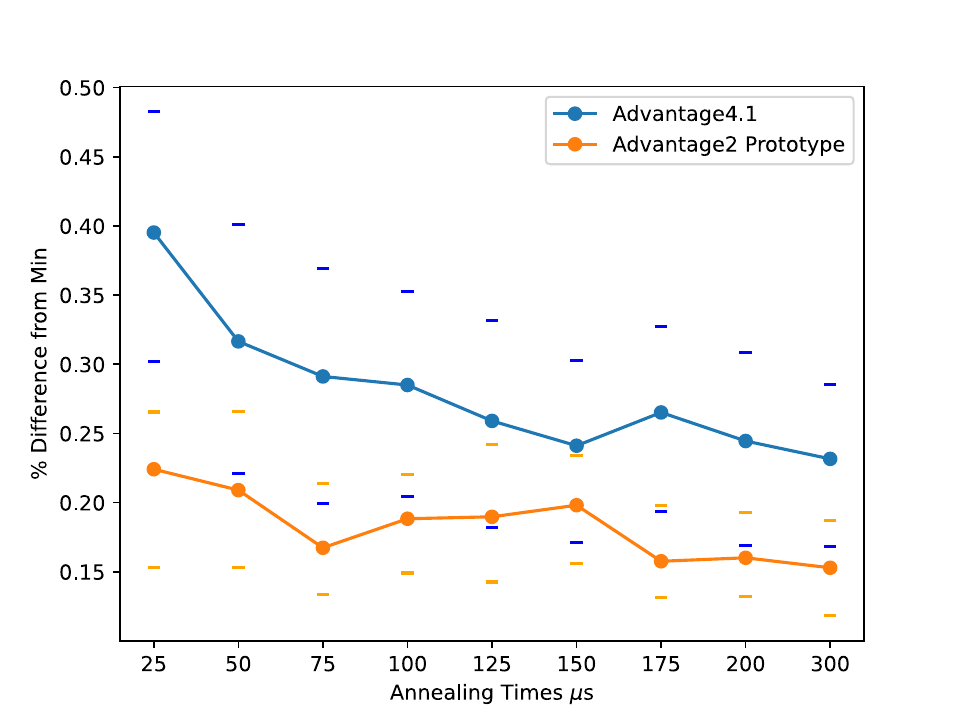}
    \caption{Comparison of QPUs using $\mathcal{F} = .5$ and the difference from the minimum energy found per annealing time.  The Advantage 2 prototype achieves more consistent results, and both display an adiabatic effect with longer annealing times.}
    \label{fig:results6}
\end{figure}

\begin{figure}[!ht]
    \centering
    \includegraphics[width = \columnwidth]{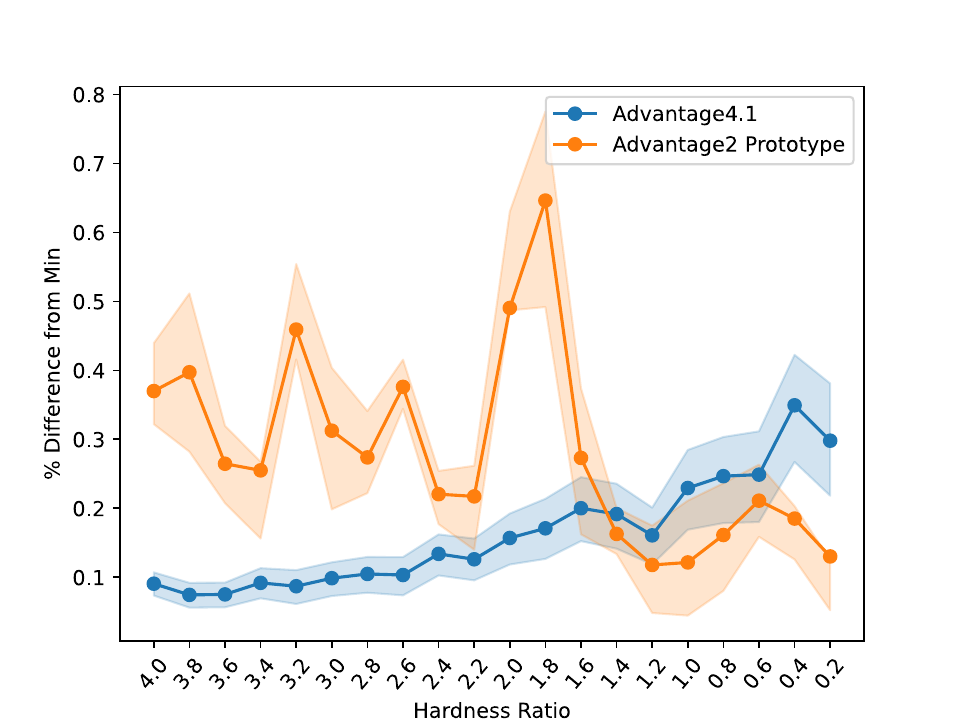}
    \caption{Comparison of QPUs across the available range of $\mathcal{F}$.  The Advantage 4.1 system became less stable as $\mathcal{F}$ decreased, while the erratic performance of the Advantage2 prototype showing increased consistency on the most difficult hardness ratios.}
    \label{fig:results7}
\end{figure}

In Fig. \ref{fig:results5} we sampled the different QPUs for hardware native Ising Models with a hardness ratio of 0.5 and for different annealing times.  On the line graph in Fig. \ref{fig:results6} we show the average solution compared to the minimum found at each annealing time, while the bars represent the first and third quartile.  From this we can see that the new Advantage2 Prototype has more consistent results, that is, more samples from the QPU are closer to the minimum found.  This increased consistency is a step forward in QPU stability and quality of results.  Both QPUs displayed adiabatic evolution with longer annealing times showing better solutions, albeit with slight deviations around 100 - 125 $\mu s$.
Fig. \ref{fig:results7} displays the same metric for different hardware native Ising problems with different hardness ratios for the same annealing time of 100 $\mu s$.  Here, an interesting trend is displayed with the Advantage 4.1 system: as the hardness ratio decreases, less consistent results are achieved with wider variance in the samples.  This intuitively makes sense, as these are the most challenging problems found using our framework, relying heavily on the entanglement between the qubits to find deeper local minima during the annealing process. 

The prototype system was clearly not as consistent, having a higher variance in the quality of results for different hardness ratios.  Still, on the most challenging problems (hardness ratio $< 1.4$), it displayed less variance compared to its predecessor and is a promising indication for better performance in the next generation of QAs.

\section{Conclusions}
In this paper we conducted Ising Model experiments on D-Wave Systems' quantum annealers with coefficients drawn from a variety of distributions and defined the RL metric, Eq. \eqref{eq:rel-dif-sa}, which served as a fair metric for evaluating the performance of quantum annealing versus simulated annealing. Furthermore, using the hardness ratio $\mathcal{F}$, we identified the inflection point that seems to separate problems amenable to simulated annealing (or potentially other classical algorithms) and problems amenable to quantum annealing. With the \KP~as a known hard-to-beat problem, we explored the performance of the Advantage 4.1 machine versus simulated annealing and provided quantitative reasoning on why we would not expect that quantum annealers outperform classical solvers in arbitrary \KP~instances (where the coefficients are drawn from random distributions) and we justified this using the hardness ratio for which \KP~is always unfavorable. Finally, we investigated the performance of the prototype machine Advantage2 versus the currently available one Advantage 4.1 and found that in the low-hardness regime, where quantum annealing performance seems to be superior to simulated annealing, the newer machine, against the older, seems to be very promising despite the lack of full implementation of new features. 

Overall, the take-home message from this paper is two-fold. The first one is that the hardness ratio analysis of a given problem may serve as a clear indicator of which technology one should focus on for that problem, and, second, the promising performance of the prototype Advantage2 machine in the regime where quantum annealers may be expected to outperform classical algorithms. \\

\vspace{1em}

\subsubsection*{Disclaimer (Deloitte)}
About Deloitte: Deloitte refers to one or more of Deloitte Touche Tohmatsu Limited, a UK private company limited by guarantee (“DTTL”), its network of member firms, and their related entities. DTTL and each of its member firms are legally separate and independent entities. DTTL (also referred to as “Deloitte Global”) does not provide services to clients. In the United States, Deloitte refers to one or more of the US member firms of DTTL, their related entities that operate using the “Deloitte” name in the United States and their respective affiliates. Certain services may not be available to attest clients under the rules and regulations of public accounting. Please see  www.deloitte.com/about to learn more about our global network of member firms.

Deloitte provides industry-leading audit, consulting, tax and advisory services to many of the world’s most admired brands, including nearly 90\% of the Fortune 500® and more than 8,500 U.S.-based private companies. At Deloitte, we strive to live our purpose of making an impact that matters by creating trust and confidence in a more equitable society. We leverage our unique blend of business acumen, command of technology, and strategic technology alliances to advise our clients across industries as they  build their future. Deloitte is proud to be part of the largest global professional services network serving our clients in the markets that are most important to them. Bringing more than 175 years of service, our network of member firms spans more than 150 countries and territories. Learn how Deloitte’s approximately 457,000 people worldwide connect for impact at  www.deloitte.com.

This publication contains general information only and Deloitte is not, by means of this [publication or presentation], rendering accounting, business, financial, investment, legal, tax, or other professional advice or services. This [publication or presentation] is not a substitute for such professional advice or services, nor should it be used as a basis for any decision or action that may affect your business. Before making any decision or taking any action that may affect your business, you should consult a qualified professional advisor.
Deloitte shall not be responsible for any loss sustained by any person who relies on this publication. Copyright © 2023 Deloitte Development LLC. All rights reserved.

\subsubsection*{Disclaimer (HSBC)}
This paper was prepared for information purposes
and is not a product of HSBC Bank Plc. or its affiliates.
Neither HSBC Bank Plc. nor any of its affiliates make
any explicit or implied representation or warranty and
none of them accept any liability in connection with
this paper, including, but limited to, the completeness,
accuracy, reliability of information contained herein and
the potential legal, compliance, tax or accounting effects
thereof. Copyright HSBC Group 2023.

\bibliographystyle{unsrt}
\bibliography{ref}

\appendix 
\section{Distribution Derivations}\label{Sec:dist-der}
For a general random variable $X$ with state space $S$, denote the mean and standard as $\mu_X$ and $\sigma_X$,  the probability distribution function (p.d.f.) as $p_X(x)$ and the cumulative distribution function (c.d.f.) as $P_X(x)$. Taking random i.i.d. variables $X_1,\ldots,X_n$ one may see that
\begin{equation}\label{eq:max-appnd}
    P\Big( \max\{X_1, \ldots, X_n \} \leq x \Big) = P_X(x)^n,
\end{equation}
\begin{equation}\label{eq:min-appnd}
 \mbox{and} \   P\Big( \min\{X_1, \ldots, X_n \} \leq x \Big) = 1- (1-P_X(x) )^n.
\end{equation}

To complete deriving closed forms for a general random variable define $\mbox{Max}=\max\{X_1,\ldots,X_n\}$ and $\mbox{Min}=\min\{X_1,\ldots,X_n\}$. Noting the inequality $\mbox{Min}<\mbox{Max}$ holds almost surely yields
\begin{equation}\label{eq:diff-deriv-appnd}
    \begin{split}
    & P\big( \mbox{Max} - \mbox{Min} \leq x \big) = P\big( \mbox{Max} \leq x + \mbox{Min} \big) \\
    & = \int_{S} P\big( \mbox{Max} \leq x + \mbox{Min} \big| \mbox{Min} = y \big)p(\mbox{Min} = y)dy 
    \end{split}
\end{equation}
To complete simplifying Equation \ref{eq:diff-deriv-appnd} we need to derive a closed forms for $P\big( \mbox{Max} \leq x + \mbox{Min} \big| \mbox{Min} = y \big)$ and $P(\mbox{Min} = y)$. From Equation \ref{eq:min-appnd} we have
\begin{equation}\label{eq:min-pdf-appnd}
p(\mbox{Min} = y) = \frac{\partial (1-P_X(x))^{n}}{\partial x} \Big|_{y}.
\end{equation}

Now for the c.d.f. $P\big( \mbox{Max} < x + \mbox{Min} \big| \mbox{Min} = y \big)$ observe we can write this distribution as $P\big( y \leq \mbox{Min},\mbox{Max} \leq x + y \big)$. Using the i.i.d. assumption we have
\begin{equation}\label{eq:intersection-appnd}
    \begin{split}
     P\big( y \leq & \mbox{Min},\mbox{Max} \leq x+y \big) = P \left( \bigcap_{i=1}^n \{y \leq X_i \leq x+y\}  \right) \\
    & = \prod_{i=1}^n P_X\left( y \leq X_i \leq x+y \right) \\
    & = \Big( P_X(X \leq x+y) - P_X(X \leq y) \Big)^n.
    \end{split}
\end{equation}
To relate Equation \ref{eq:intersection-appnd} back to Equation \ref{eq:diff-deriv-appnd} we calculate the p.d.f. by differentiating with respect to general variables $x_1$ and $x_2$ so that 
\begin{equation}\label{eq:diff-pdf-appnd}
p(y_1,y_2) :=  \frac{\partial^2 \Big( P_X(X \leq x_2) - P_X(X \leq x_1 ) \Big)^n \mathbf{1}_{x_1<x_2} }{\partial_{x_2} \partial_{x_1}}  \bigg|_{y_1, y_2},
\end{equation}
where $\mathbf{1}_{x_1<x_2}$ is the indicator function that is $1$ when the inequality holds and $0$ otherwise. 

Taking Equation \ref{eq:min-pdf-appnd} and Equation \ref{eq:diff-pdf-appnd}, Equation \ref{eq:diff-deriv-appnd} has the form
\begin{equation}\label{eq:diff-deriv-FINAL-appnd}
   P\big( \mbox{Max} - \mbox{Min} \leq x \big) = \int_{S} \int_{y}^{y+x} p(y,z) p_{min}(y) dz dy.
\end{equation}

Finally, it is well-known that for independent random variables $A$ and $B$, and $Z:=AB$ then the p.d.f. of $Z$ is given by the convolution
\begin{equation}\label{eq:two-vars-appnd}
    p_Z(z) = \int_{S} p_A(x)p_B(z/x)\frac{1}{|x|}dx.
\end{equation}
One may see this equality since, from independence, we have
\begin{equation*}
\begin{split}
    & P_Z(Z \leq z) = P_Z(B \leq z/A) = \int_{S} P(B \leq z/x)p_A(x)dx \\
    & = \int_{S} \int_{S} \frac{p_{B}(z/x)}{|x|} p_A(x)dxdz
\end{split}
\end{equation*}

For an overview of the information in the appendix see Billingsley \cite{billingsley2017probability}.

\end{document}